# Heat-flash travel just above a deep Mediterranean seafloor

by Hans van Haren

Royal Netherlands Institute for Sea Research (NIOZ), P.O. Box 59, 1790 AB Den Burg, the Netherlands.
e-mail: hans.van.haren@nioz.nl

**Abstract.** The deep sea is weakly stratified in density but shows considerable variations in turbulent motions in all three directions. When registered by moored high-resolution temperature 'T'-sensors, the motions cause variations of 0.01°C or less and in time of minutes or less, which is much faster than hours or longer of internal waves. Occasionally, T-sensors close to the seafloor register minute-long flashes of 0.0005-0.001°C warmer than the environment. When singular, such flashes may be artefacts. However, in a large mooring-array with 45 vertical lines at 9.5-m horizontal distances, near-seafloor heat flashes are seen to travel, most likely with internal-wave instabilities in overlying stratified waters. The instabilities seem to release the flashes from a geothermally heated seafloor of which turbulence convection is suppressed by warmer waters from above. The forms and turbulence intensity of these rare signals are compared with those induced by a Remotely Operated Vehicle working near the array. Other causes like unidentified marine mammal passing are hypothesized.

# 1 Introduction

As with all observational methods, instrumental knowledge is required to distinguish between artificial and 'real-signal' data. The distinction not only separates instrumental noise from environmental signals, but also time-varying electronic drifts or bias from waves, and digitization glitches from short-term turbulent overturns. In a laboratory set-up, instruments may be regularly tuned during an experiment for better signal compared with artificial data. Such tuning is not possible for remotely-operating stand-alone oceanographic instrumentation on deep-sea underwater moorings that release their data after instrument recovery. Data from those oceanographic instruments are subject to elaborate post-processing, involving separation of known artificial and environmental signals, but especially also of unknown.

Improved instrumentation such as high-resolution temperature 'T-'sensors mounted on a mooring-array of multiple closely spaced vertical lines deployed in weakly stratified waters may provide new insights of propagation of waves and various turbulent mixing processes besides rare passages of warm water blobs by deep-sea animals or remotely operated vehicles 'ROVs'.

In this paper, several of such rare warm-water passages are presented and discussed in data from the deep Western Mediterranean Sea, where most of the dynamical and artificial temperature variations are in the range of 0.0001-0.01°C. The focus will be on 'heat flashes', temperature variations of 0.0002-0.001°C above environmental values that last 0.0005-0.005 day (45-450 s). Such flashes are mainly observed within a few meters above the seafloor.

# 2 Materials and Methods

A volume of nearly half-a-million cubic meters of seawater has been sampled using 2925 self-contained high-resolution 'NIOZ4' T-sensors. The ensemble 'large-ring' mooring was deployed in drag-parachute controlled free-fall at the <1° flat and 2458-m deep seafloor of 42° 49.50′N, 006° 11.78′E and was underwater between October 2020 and March 2024. The site was near the neutrino telescope KM3NeT/ORCA (Adrián-Martinez et al., 2016) and just 10 km south of the steep continental slope of the Northwestern Mediterranean Sea.

The large-ring mooring had a diameter of about 70 m (Fig. 1). The eighteen 12-m long and 0.61-m diameter steel pipes held a steel-cable grid for rigidness, like spokes in a wheel. Perpendicular cables were 9.5 m apart. At cable intersects, 2.5-m diameter 'small' rings were mounted that each held a 125-m long mooring line with 65 T-sensors below a single 1.45-kN buoy. After unrolling, the T-sensors had their sensor-tip directed down along their vertical line to avoid accumulation of sediment. Of eight small rings, imaginary intersects were at the steel pipes, so that special off-set mounting was needed with three assist cables (van Haren et al., 2021). Upon landing at the seafloor, the orientation of the ring was directed to the NNW, pointing at 337 °N. The 45 buoys lifted the cable grid in a dome with its center at a height of $h = 2.0\pm0.2$ m above seafloor (Appendix A). On vertical lines near the edges of the cable grid, close to the steel pipes, the lowest T-sensor was at about $h = 0.7\pm0.2$ m.

Fig. 1 shows the numbering of the 45 vertical mooring lines, which were ordered in six groups for synchronisation purposes. As with previous NIOZ4 T-sensors (van Haren, 2018), the individual clocks were synchronised via induction to a single standard clock on the mooring-array every 4 hours, so that all sensors were sampled within 0.01 s. The single synchroniser 'S' was located at the small ring of central line 5.1, line 1 of synchronization group 5. (Henceforth in the text vertical lines will be indicated without period). Three buoys also held an AquaDopp single-point acoustic current meter.

Per vertical line, two T-sensors registered acceleration-tilt besides temperature. These sensors were located just above the lowest and just below the uppermost T-sensors. The other 63 sensors only registered temperature and were mounted at 2.0-m intervals. Unintentionally, all T-sensors switched off when the file size on the 8-GB Kingston memory card reached 30 MB, which was likely due to a formatting error. It implied that a maximum of 20 months of data was obtained for the 63 temperature-only sensors, and 5.5 months for tilt-temperature sensors. All recorded data at an interval of once per 2 s. Of the total of 2835 temperature-only sensors, between 50 and 150 data records were interpolated between neighbouring sensors because of some instrumental failure or bias, depending on moment in the record.

As detailed elsewhere (van Haren, 2018), laboratory-bath calibration yielded a relative precision of <0.001°C. Post-processing of data involved correction for instrumental electronic drift of about 0.001°C

mo$^{-1}$ after aging. This correction was established by referencing daily averaged vertical profiles, which must be stable from a turbulent-overturning perspective in a stratified environment, to a smooth polynomial without instabilities. The multiple lines were referenced to an arbitrary single mean value, per daily period. In addition, because vertical temperature (density) gradients are so small in the deep Mediterranean, so that buoyancy frequency N = O(f) where f denotes the inertial frequency, reference was made to periods of typically one hour duration that were quasi homogeneous with temperature variation smaller than instrumental noise level (van Haren, 2022). Such >124-m tall quasi-homogeneous periods existed on days 350, 453, and 657 (the latter two -366 in 2021) in the records. This double drift correction was necessary for quantification of turbulence using the method of Thorpe (1977) under weakly stratified conditions and typical energy-containing scales between 1 and 100 m. The T-sensors at the 45 lines of the large-ring mooring allowed for improved statistics of turbulence values, besides creation of small movies for investigation of propagation of phenomena like waves, eddies, turbulent clouds (for an overview see van Haren et al., 2026), and rare heat flashes.

## 3 Results

The focus is on the first full year of observations, between days 306 and 671, with some emphasis on the first months when tilt-temperature sensors were still recording.

### 3.1 Rare multiple line travel of warm-water flashes

On day 436.87 (late evening of Friday 12 March 2021), small heat flashes of up to 0.001°C above environmental values were observed in the lowest T-sensor above the seafloor. Initially these were seen at lines 66, 64, 65, and 67 in that order, followed by very weak wavy variations of about 0.00003°C. A quasi-3D movie of 3000-cpd (short for cycles per day) low-pass filtered 'lpf' temperature records demonstrates that the sequence continued in weaker form to lines 12-16 (Fig. 2). Higher-up in the water, slow cat's eye shrinking and expanding of parametrically induced quasi-mode-two instability 'waves' are discernible, with associated turbulence generation. Duration of the passage of heat flashes varied between 0.0005 and 0.01 days (45 and 900 s) at different sensors. The average propagation was in E direction (cf., Fig. 1) at a speed of 0.06 m s$^{-1}$, which was about twice the average particle velocity u in ESE direction measured at h = 126 m (Fig. 2 small right panel). For reference in the 3D cube of the movie, white numbers are indicated when the line's lowest T-sensor becomes warmer than the environment. The upper panel of the 72-s (1440-s real-time) movie shows the lower 80 m above seafloor time-depth series from line 16. It includes heat flashes near the seafloor, e.g., at the white vertical line in the upper image of Fig. 2, and a sequence of 'cat's eye' in- and decreasing of isotherms higher up. The variations in isotherms are visible in the movie as gentle shrinking and expanding of colours along vertical lines in the 3D cube. As will be elaborated below, the associated local vertical mode-2 motions are a non-negligible means for internal wave-induced turbulence across the (weak) stratification, and may also have effect on the seafloor.

In unfiltered, but de-trended with common mean, time series of the lower four T-sensors, including tilt-temperature sensors, the duration and form of heat flashes is well visible (Fig. 3). Flashes are mostly found at the lowest T-sensor, of which the nominal but not the exact heights are given. With respect to given heights, lines 16 and 67 differ by -0.7 m, line 66 by -0.5 m, and the others by 0 or +0.5 m, depending on distance to the nearest steel pipe (cf., Appendix A). In none of the lines, flashes are observed at nominally h = 5.5 m, a few at h = 3.5 m, while in half the cases at h = 2 m.

The warmest flashes are observed in the NW of the ring, line 66. Intense flashes have a typical duration of 0.001 day (about 90 s), but shorter duration ones are also visible. Given that all flashes are observed at the lowest T-sensor per line, with largest value at line 66, it is anticipated that their source lies near the seafloor to the NW of the large ring. As for the size, initially flashes are seen at separate lines only, so the horizontal scale is <9.5 m (and <3-m vertically). Its elongated extent at line 65 and back-bending occurrence at line 67 either demonstrate a whirling motion advecting the flashes, or an expansion of warm area of complex form. The amplitude increase at line 16 following the steady decrease along the sequence is probably due to the relatively short height above seafloor of sensors close to a steel pipe.

Assuming that the steady decrease in amplitude reflects a single source and transport advection by local waterflow, the somewhat irregular order of appearance at lines, in both northward and eastward directions, complicates the interpretation. As for the source, a single vent-like geothermal warming is possible, but no external vent-tubes were observed in the vicinity and conditions for observing general geothermal heating occurred about 60% of time, throughout the record. This contrasts with the rare sequence in Figs 2, 3, which only somewhat compares with another event presented below, during the one year of observations. Presumably, an unidentified external heat source visited the site, for example a deep-diving submarine or a sperm whale, perhaps urinating. Before elaborating on these hypotheses below, a comparison is made with mid-height physical turbulence events appearing around the same time.

**3.2 Mid-height occurrence of internal-wave turbulence**

Four records of neighbouring mid-height, around h = 55 m above seafloor, T-sensors on the lines of Fig. 3 demonstrate also a sequence of about 0.0003°C temperature variations of typically 0.003-day (250-s) duration and occurring about every 0.01 day (Fig. 4). When temperature falls at some sensors, it rises at others around the same time, which evidences a local mode-two motion. (In the lower-left panel, this is in opposite to the upper panel of Fig. 2: where isotherms separate, temperature records are about the same and turbulent overturning is expected, e.g., on day 436.882; where isotherms are close together, temperatures differ and enhanced stratification reduces turbulence scales, e.g., around day 436.886). Their oblique appearance deforms from exactly mode-two, and is more akin to turbulent overturning, probably under larger-scale shear (Thorpe, 2005).

The occurrence interval and size of these quasi-mode-two motions is faster and shorter than the smallest freely propagating internal waves, which amounts about 0.14 day for given small-scale stratification, as far as can be established from 2-m vertical resolution. The discrepancy in duration by a factor of 14 is about three times longer than found in laboratory experiments (Davis and Acrivos, 1967; Thorpe, 2005), if the smallest freely propagating internal waves generate the 0.01-day long mode-two motions as parametric instabilities. (For comparison, a heat flash of 0.001°C would create hypothetical stratification equivalent to small-1-m-scale buoyancy frequency of $N_1 \approx 16f$, or a free internal-wave periodicity of about 0.04 day, which approaches the laboratory factor of 4.) The sequence between the lines is different from that presented in Fig. 3, and more steadily propagates in E direction at a speed of $c = 0.10$ m s$^{-1} \approx 4u$, u measured at h = 126 m. It is thought that growing parametric instabilities may break and thereby disintegrate their generating progressive primary internal waves (Davis and Acrivos, 1967), by which they are expected to be transported. Their amount of turbulence is calculated in Section 3. Although the quasi-mode-two motions have the same order of duration as the near-seafloor heat flashes, they do not directly seem to associate with them, because they do not occur at precisely the same time and slightly differently between different lines. Their potential relationship is investigated below in an enlargement of filtered data.

Fig. 5 is the lpf version of temperature from two lower and two mid-height T-sensors, extended over a period of 0.2 day. Mean temperature values are referenced to a smooth polynomial over the entire 124-m vertical range, for drift correction. The near-seafloor heat flashes (in red graphs) are seen embedded by a weak local mode-two variation in temperature, peaking with a local low a few meters above (green graphs). They occur when mid-height mode-two motions are at reduced variation, i.e. increased isotherm straining, followed by a drop in temperature at lower mid-height (magenta graphs). The observation of mode-two variations and associated turbulent overturning across the vertical range of moored T-sensors suggests some relationship between internal-wave induced mid-height turbulence and that occurring just above seafloor. The internal-wave variations possibly trigger/release geothermal heating to convect upwards during lower mid-height cooling phase, and seafloor convection is depressed during lower mid-height warming phase. Besides the heat flashes, geothermal-heat convection is rather limited during this period.

**3.3 Another sequence, but different**

The near-seafloor heat flashes in Figs 2, 3 are thus rare that only one other event has been detected in the year-long records. This other event is shown in Fig. 6, upper-left column. In contrast with conditions of reasonably stratified waters for Figs 2, 3, this event is observed under very weakly stratified conditions at only three lines, starting within the large ring at line 11 and propagating in WNW direction passing lines 65 and 67 at a mean speed of 0.05 m s$^{-1}$. During this period at h = 126 m, waterflow equals u = 0.005 m s$^{-1}$, negligibly small in SW direction. While spikiness and duration of the flashes resemble those of Fig. 3, it is puzzling why the record from the temperature-tilt sensor starts with a spike and the flash is extended to twice the common duration at the lowest sensor of line 67. The flashes occur about halfway during an episode of near-homogeneous waters, when a 1-h long 0.0001-0.0002°C elevation in temperature is observed at upper T-sensors compared to those closer to the seafloor. Despite favourable conditions of very weakly stratified waters, no convection turbulence due to general geothermal heating was observed during this period.

**3.4 Isolated spikes at the lowest T-sensor**

Seldom, albeit occurring more often than sequences like depicted in Figs 2-5, singular heat flashes are observed at the lowest T-sensor of some vertical lines, e.g., line 25 (Fig. 6 lower left). Given that the flashes are observed at no other lines during or close to the time of appearance, no direction and propagation speed can be established. For the given example, conditions are near-homogeneous with a vertical temperature difference between uppermost and lowest T-sensor of $\Delta\Theta < 0.0001$°C. Despite the small vertical temperature variations barely extending above noise level, temporal variations are visible in the duration of the flashes, being about 0.003 day (250 s), and oscillations vary between 0.001 and 0.005 days, besides > 0.04 d longer duration ones (not shown). Moreover, vertical mode-two is discernible locally with next T-sensors up, e.g. a weak depression at 2 m during the heat flash at 1.5 m, and across the entire 124-m vertical range after day 370.67.

**3.5 The temperature signature of an ROV**

While the above examples suggest a correspondence between heat flashes, possibly geothermal heating from below and internal wave turbulence from above, a comparison should be made with temperature signals from known other effects. On day 323, ROV 'Victor' (Ifremer, La Seyne-sur-mer, France) approached the large-ring mooring. Its purpose was to set free the acoustically released but mechanically stuck rope of the large-ring deployment drag parachute. The stuck rope, one out of six released, had scraped several T-sensors from nearest vertical line 18. During operations also several fallen-off T-sensors were collected outside the large ring from the seafloor. The ROV's engine heats the local waters and its impellors create additional local turbulence and advective waterflow. The ROV always stayed outside the large ring, but sometimes rested on a steel tube, e.g. on day 323.675.

While the ROV's temperature signals are > 0.001°C above environmental values at some T-sensors of line 18, they rapidly decrease away from the 'source', with the nearest three lines showing weaker and shorter duration artificial temperature elevations at their lowest three T-sensors mainly (Fig. 6 upper-right column). Weaker signals are found towards the south (lines 22-25). An approximately 10-m high 'cloud' of warmer water is observed on day 323.655 around h = 25 m (line 16) moving up to h = 50 m (line 26) whilst decreasing in intensity and duration. This cloud is not seen northward, also not at line 18, and may reflect the expulsion by the impellors. No artificial temperature elevation is detected further to the West at lines 11, 12, 17, and 21. The mean waterflow was 0.04 m s$^{-1}$ in ENE direction, measured at h = 126 m, and perpendicular to the direction of the cloud. Environmental conditions were near homogeneous, with a positive vertical temperature difference of about 0.0001°C over 124 m.

The type of temperature signals is akin to those of others presented in Fig. 6. The ROV's warm signal remains near the seafloor, also > 9.5 m away from the source, and is short-lived with flash durations between <0.001 and 0.005 days. Irregularly, temperature is not highest at the seafloor, but at a T-sensor above, e.g. on lines 15 and 16. Considering the likelihood of the source near line 18, the with 2.8-m high 0.5 m above the seafloor at the nearby steel pipe, the dispersal is about 0.035 m s$^{-1}$ in SW direction, which is against the flow direction measured at h = 126 m. No increase in the vertical is observed across the approximately 10-m horizontal distance, despite the expected free convection turbulence occurring by the warm water. Apparently the turbulence is a slow process.

**3.6 Geothermal-heating signals at lower T-sensors**

Heat flashes are observed more often uniquely at lowermost T-sensors of generally lower-lying 'corner-lines' like 17 and lines like 25, under weakly but stratified-water conditions and less so under near-homogeneous conditions when the 124-m vertical temperature difference equals < 0.0002°C. Fig. 7 shows occurrence of flashes that are visually detected at the lowermost sensor of line 17, which did not register the sequence of Figs 2, 3. Despite most occurrence of heat flashes under (weak) stratified-water conditions, the flashes are associated with geothermal heating from below but which is suppressed by the warmer waters above. This may explain the roughly 20-30-day periodicity of the heat flashes, and their lack of appearance higher up. While the 20-30-day periodicity may thereby be associated with mesoscale variability governing the variation in stratification conditions, the flashes' signature is verified before the ROV collected fallen-off T-sensors from the seafloor on day 323. A T-sensor lying on the seafloor has its measuring tip within 0.01 m from the sediment.

On day 317, inertial-period temporal variation in temperature was observed throughout the range of observations (van Haren, 2023). Despite the very weakly stratified conditions, with $\Delta\Theta \approx 0.0001°C$ over $h = 124$ m, heat flashes alternate with time. They are suppressed and remain in the lower $h = 2$ m above seafloor (e.g., Fig. 6 lower right), before becoming released in geothermal-induced flares that may reach several tens of meters up from the seafloor and which are of longer duration than heat flashes lasting hours. During suppression by warming from above, the record of a fallen-off T-sensor shows rapid variability with temperature variations akin to those induced by the ROV (Fig. 6 upper-right column). In case of heat release into overlying waters, near-seafloor temperature variations decrease as excess heat is distributed over a larger vertical range. Apparently, even under weakly stratified conditions suppression of convection turbulence is so strong that rarely geothermally-heated temperature peaks, heat flashes, reach above $h > 1$ m, as observed.

### 3.7 Turbulence effects of flashes

As all observed heat flashes are characterized by elevated temperatures above environmental values near the seafloor, they are unstable and may provide a contribution to turbulence values. On the other hand, their duration is generally short. Their effects on general turbulence values are verified below using the method proposed by Thorpe (1977).

Concerning the passage of flashes on day 436, small spikes are seen in the vertically averaged dissipation rate values for line 16 (Fig. 8a, b). Only the spikes around days 436.848 and 436.883 are associated with heat flashes near the seafloor, those after day 436.9 are not involved. Thus, the contribution of these flashes is negligible to the overall-mean turbulence, which is dominated by parametric instabilities following internal-wave action from above. The 0.2-day and 124-m time/depth mean values for this period under 'stratified-water' conditions are: for turbulence dissipation rate $<[\varepsilon]> = 7\pm3\times10^{-11}$ m$^2$ s$^{-3}$, vertical eddy diffusivity $<[K_z]> = 6\pm2\times10^{-4}$ m$^2$ s$^{-1}$ under large-scale buoyancy frequency N = <[$N_s$]> = 1.5±0.4×10$^{-4}$ s$^{-1}$ = 1.1f. These turbulence values are comparable with well-stratified open-ocean turbulence values (e.g., Gregg, 1989; Polzin et al., 1997; Yasuda et al., 2021) and about half of well-developed convection turbulence induced by general geothermal heating in the present area: $\varepsilon_{GH}$ = 1.2×10$^{-10}$ m$^2$ s$^{-3}$ (van Haren, 202b submitted) which corresponds with the local average mean heat flux of 0.11 W m$^{-2}$ determined from geophysics measurements (Pasquale et al., 1996). Calculations for neighbouring lines demonstrate that the heat flashes contribute <1% to mean turbulence values, in this example.

For comparison, during ROV working on day 323 (Fig. 8c, d), the 0.2-day and 124-m time/depth mean values for line 15 during this 'very weakly stratified' period are: <[$\varepsilon$]> = 3±1×10$^{-10}$ m$^2$ s$^{-3}$, <[$K_z$]> = 4±1.5×10$^{-3}$ m$^2$ s$^{-1}$ under N = <[$N_s$]> = 7.5±2×10$^{-5}$ s$^{-1}$. Thus, under four times smaller mean stratification, turbulence values are approximately four times larger than the previous example. However, the time series of vertically averaged dissipation rate values demonstrates about the same level as in Fig. 8b, except during two periods of spiking values around days 323.61 and 323.67 when turbulence dissipation rate values reach 10$^{-8}$ m$^2$ s$^{-3}$. In the calculations, the secondary drift correction is not applied, for comparison with Fig. 8a, b. The mean turbulence values in Fig. 8b, d are three times the values from lines to the West in the large-ring, but about 30 times smaller than those of line 16 nearest to the working ROV (apart from damaged line 18). As a result, artificial heat flashes can dominate turbulence calculations under very weakly stratified conditions.

**4 Discussion**

As not one cold flash has been detected in the inspected year of observations, and given the more prominent occurrence near the seafloor, the source of observed heat flashes seems evident. Either they result from an artificial or biological source, or from geothermal heating. In addition to the source, a means to travel may be related to the source or be independent. For example, an ROV has impellors that can drive a waterflow advecting the engine's excessive heat. Or, mesoscale eddies, internal waves and turbulence motions operating some distance above the seafloor can move geothermal heat originating at the seafloor.

Further investigation is required on the release of general geothermal heat from the seafloor. Apparently, excess warmer waters stay in the lower 2 m from the seafloor most of the time, becoming only convected vertically in 10-100 m high flares when conditions are favourable and stratification is negligible so that $N < 0.5f$. The convection-turbulence flares are larger than heat flashes in spatial extent and in time, but they exceed environmental temperature values only by values $O(0.0001)°C$. Any very weak stratification, even $N = 0.75\text{-}1f$, is sufficient to suppress the heated-seafloor waters from convecting, whereby some accumulation of up to $0.001°C$ at $h = 1.5$ m occurs. Apparently, excess heat is absorbed by passing water flows, and occasional heat flashes are released at lowest (corner-line) T-sensors.

Although it cannot be ruled out that the heat flashes are transported via internal waves' associated turbulence following parametric instabilities several tens of meters from the seafloor, it remains surprising that flashes are a rare phenomenon at instrumentation only 1.5-2 m above seafloor. The fallen-off T-sensor with its sensor-tip at $h = 0.01$ m revealed much more often, albeit not continuously and highly variable, temperature variations of equivalent value. However, the variations never exceeded a difference of $0.001°C$ compared to nearby values. The geothermal-heat source thus seems variable as well, like the response in a pot over a uniformly heated plate. Geologically, geothermal-heat sources may also vary over short distances $O(100)$ m (Kunath et al., 2021), or perhaps smaller.

As for unidentified sources, the near-seafloor heat flashes all have a similar appearance. No clear identification can be given for ROV-induced or geothermally-induced flashes, other than distance from the seafloor and distance from the source. If the travelling flash on day 436 is due to ROV, or perhaps a whale, they have stayed some 10 m from the nearest vertical line, outside the large ring. As no planned sea operations were ongoing on that day near the large ring, a curious whale may have urinated, or a geothermal flash was released by some favourable wave passing above. Whatever, the flash propagated in the direction of the wave and mean flow higher-up.

Although the statistically unstable flashes do not pose a large problem for calculation of turbulence values from moored high-resolution temperature sensors, except when an ROV is operating nearby, their detection challenges the instrumentation and post-processing. Considering their duration and occurrence over multiple independent T-sensors, they are not electronic 'glitches' or 'spikes', which also occur (quite rarely) and commonly consist of a few data point (< 10-s duration). Detailed data-quality control search thus also provides insight in unknown phenomena.

**5 Conclusion**

A rare sequence of heat flashes is observed traveling through the northern part of a 70-m diameter large-ring mooring about 2-2.5 m above flat seafloor at 2458 m in the Northwestern Mediterranean. In an environment where the vertical density stratification is very weak, the buoyancy frequency equals the inertial frequency and all dynamics are contained within temperature variations <0.01°C, the flashes have maximum excess temperature of 0.001°C above surrounding values and last between 1 and 7 minutes. They appear as spikes as if due to electronic failure, which they are not. They travel with internal-wave instabilities in overlying water up to some 50 m above seafloor. This suggests an unexpected mechanism of such interior instabilities releasing flashes from the geothermally-heated seafloor. Turbulent convection is otherwise suppressed by the weakly-stable stratification. Possible unidentified source is a unique visit from either a manned or unmanned submarine, or a deep diving whale.

*Data availability.* Only raw data are stored from the T-sensor mooring-array. Analyses proceed via extensive post-processing, including manual checks, which are adapted to the specific analysis task. Because of the complex processing the raw data are not made publicly accessible. Current meter data are available from van Haren (2025): "Large-ring mooring current meter and CTD data", Mendeley Data, V1, https://doi.org/10.17632/f8kfwcvtdn.1. The movie to Fig. 2 can be found in van Haren, Hans (2026), "Movie to: Heat-flash travel just above a deep Mediterranean seafloor", Mendeley Data, V1, https://doi.org/10.17632/mdw59fv2vv.1.

*Competing interests.* The author has no competing interests.

*Acknowledgments.* This research was supported in part by NWO, the Netherlands organization for the advancement of science. Captains and crews of R/V Pelagia are thanked for the very pleasant cooperation. The team of ROV Victor performed an excellent underwater mission to set free the drag parachute of the large ring. NIOZ colleagues notably from the NMF department are especially thanked for their indispensable contributions during the long preparatory and construction phases to make this unique sea-operation successful. The KM3NeT Collaboration built, operated and maintained the KM3NeT research infrastructure, and I am grateful being part of it. For suggestions on visualization of 3D-motions I am indebted to P. Coyle, R. Bakker, J. van Bennekom and E. Keijzer.

**Appendix A Doming of steel-cable grid**

In contrast with conventionally anchored single-line moorings, the multiple-line large-ring mooring has a flexible steel-cable grid attached to grounding steel pipes (Fig. 1). Although also common single-line moorings stretch under buoyancy, thereby displacing instrumentation upward from their nominal positions, the large-ring mooring can further stretch because of its steel-cable grid. Based on cable information, and on-quay cable stretching and tensioning, several mathematical models were discussed by van Haren (2026, submitted). The models were verified (Fig. A1) using physical information during a passage of highly turbulent and well-stratified warm waters.

The cable grid was attached to the center of the 0.61-m diameter steel pipes. The pipes sank $0.07\pm0.02$ m in the sediment (van Haren et al., 2021), so that the cable attachments were at h = $0.24\pm0.02$ m above seafloor. For all cable cross-sections a near-parabola was established. Consistent statistical significance was found to within $\pm0.2$ m, see the error bar.

While the observed doming is close to the mathematical parabola models, larger height-correction values are observed in the center, with slightly steeper grid cables that approximately obey the maximum 5°-slopes as found during on-quay tensioning. Corrections to vertical positioning of T-sensors are therefore feasible and necessary, because the difference between the center and the edges of the grid is approximately $1.5\pm0.2$ m. The small height difference may have noticeable effects on stratification-suppressed geothermal heating and on the development of convection turbulence, or lack thereof.

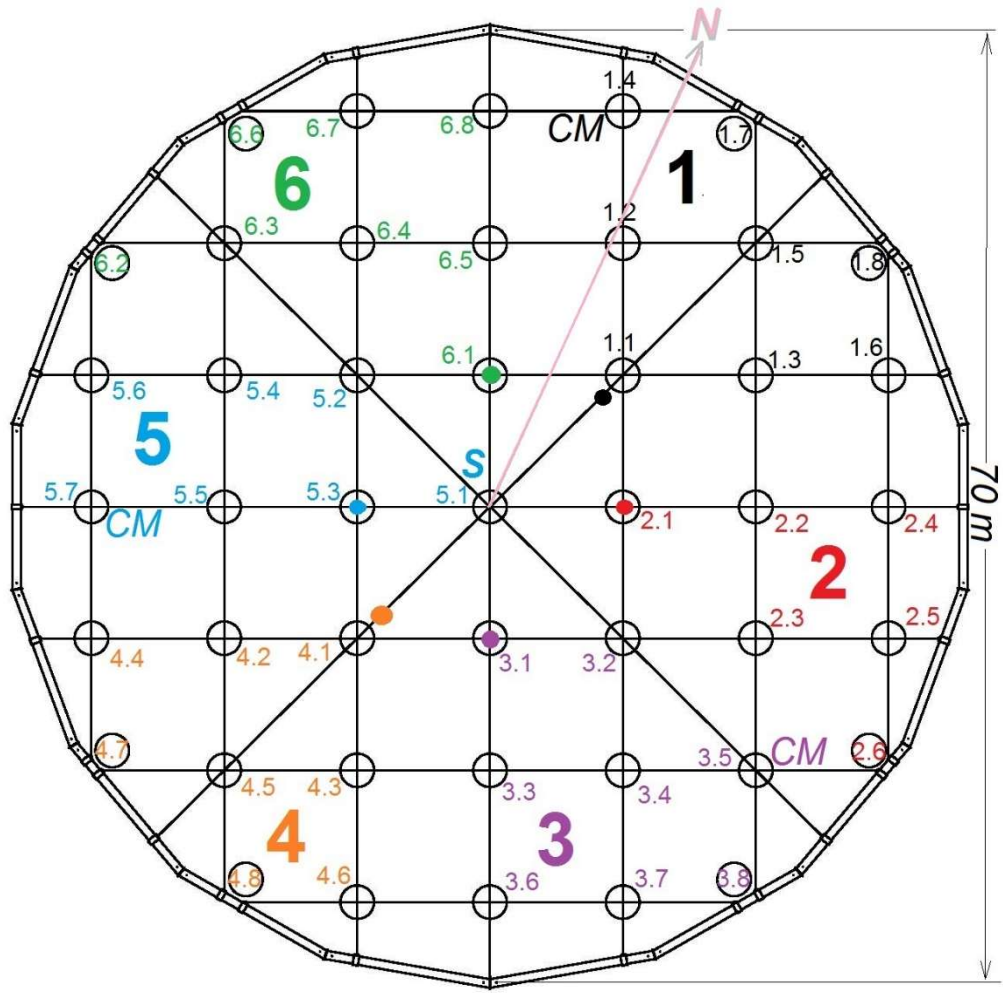

**Figure 1.** Plan-view of 'large-ring mooring' layout, including steel-cable grid and, at cable intersections, 2.5-m diameter small rings originally holding 125-m long mooring lines with 65 temperature 'T-' sensors and a single buoy each. The vertical mooring lines are numbered in six synchronisation groups with colour dots indicating group nodes, and synchroniser 'S' at line 51. Here and elsewhere in the text, lines are indicated without period for short. Lines 14, 35 and 57 held a current meter 'CM' at the buoy. Eight 'corner-'lines 17, 18, 26, 38, 47, 48, 62, and 66 were not at cable-grid intersections (van Haren et al., 2021). The light-pink arrow orients to geographical North.

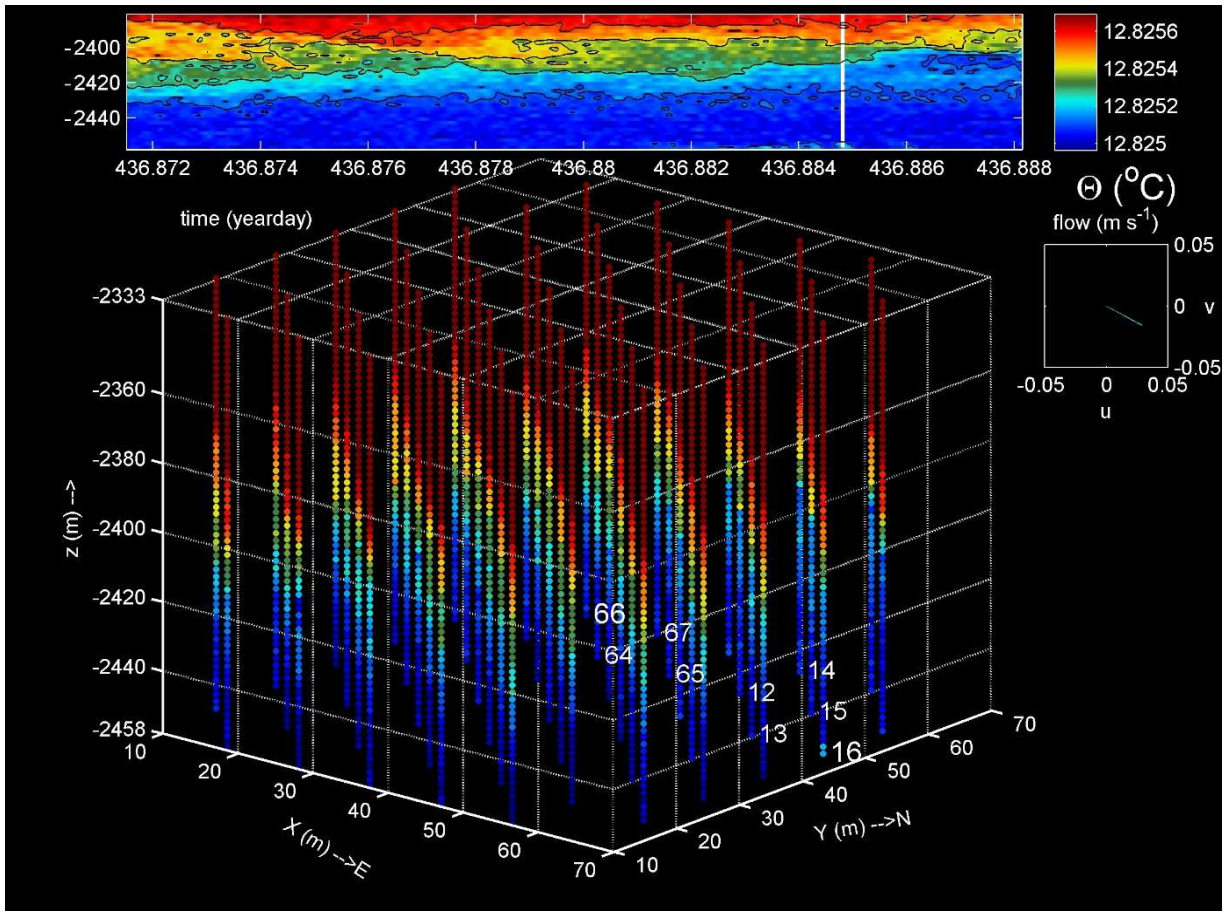

**Figure 2.** Quasi-3D movie from low-pass filtered 'lpf' temperature data of about 2800 T-sensors in nearly 0.5-hm$^3$ mooring-array. Each sensor is represented by a small filled circle, of which the colour represents a Conservative Temperature (IOC et al., 2010) in the scale above (entire range: 0.0007°C). In the movie, above the cube, which is vertically depressed by a factor of about two, a white time-line progresses in a 1440-s/124-m time/depth image from line 16 on the east-side of the cube. The 72-s movie is accelerated by a factor of 20 with respect to real-time. In the small panel to the right in this figure, but not appearing in the movie, the mean flow is indicated, measured at 126 m above seafloor.

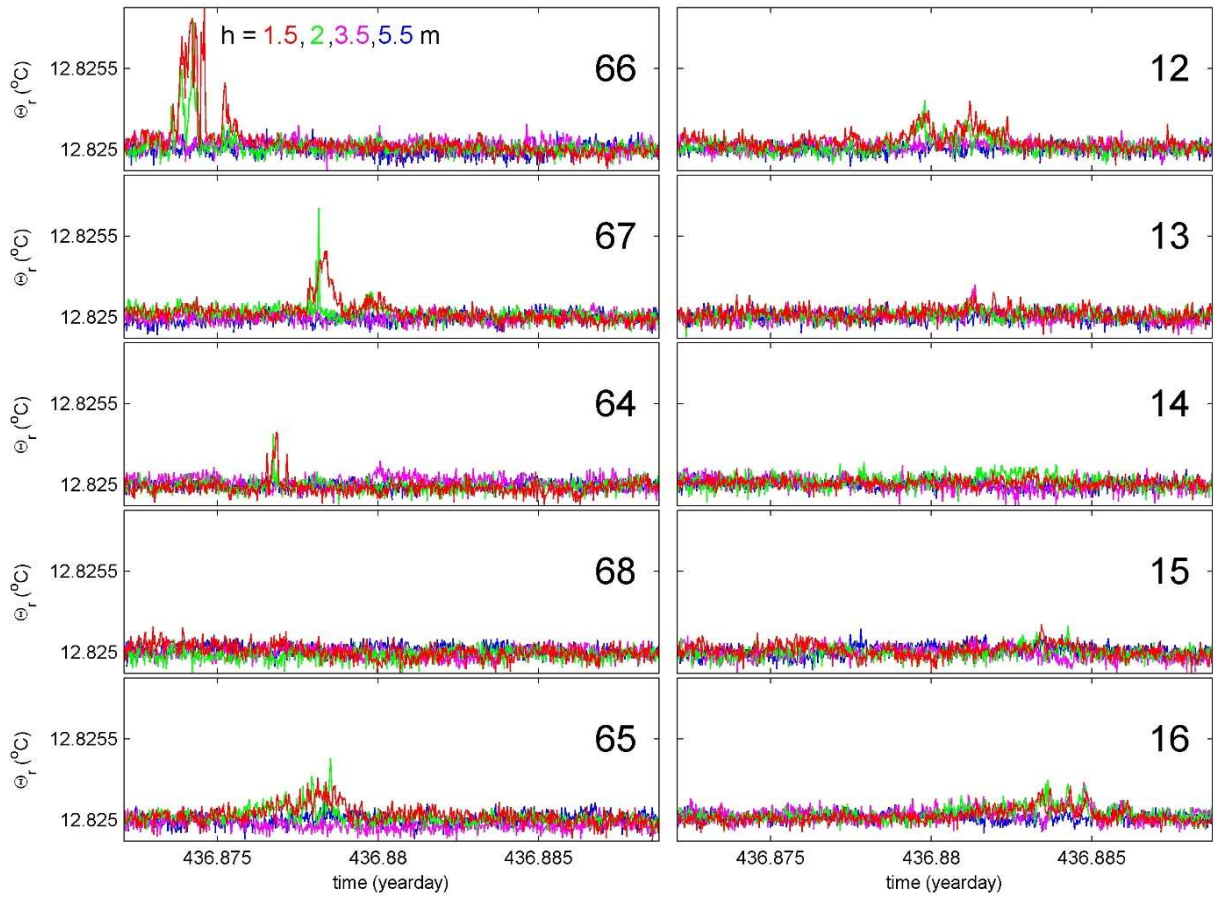

**Figure 3.** Heat-flash travel registered at 9 (out of 10 plotted) lines by the lower 2 (out of 4 plotted) T-sensors for the 1440-s real-time period of Fig. 2. The four unfiltered, detrended, and relative to common-mean Conservative Temperature time series are for heights h just above seafloor as indicated in the upper-left panel.

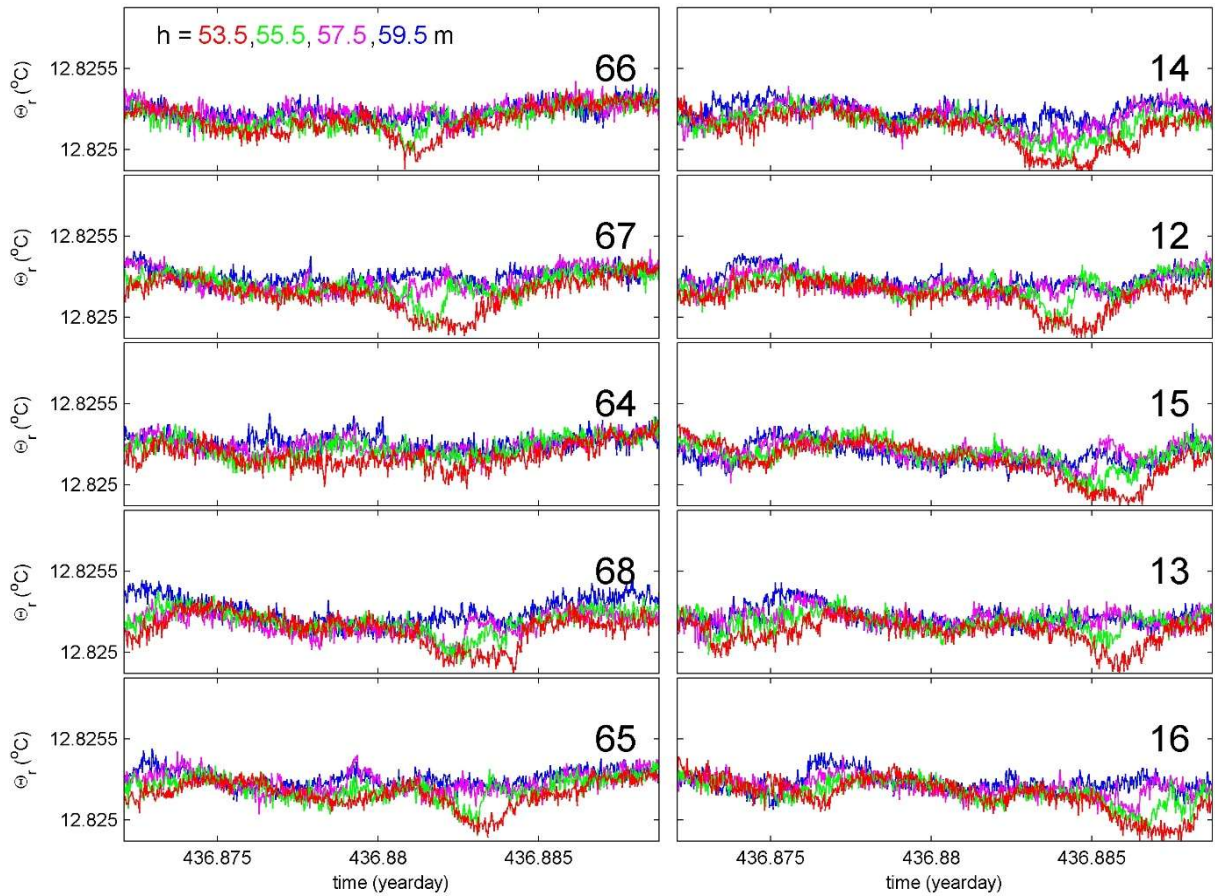

**Figure 4.** As Fig. 3, but for mid-heights showing no heat flashes but small-scale overturns that travel in the same E direction as flashes of Fig. 3, but at twice the speed, i.e. four times the waterflow (particle) speed in ESE direction measured at h = 126 m.

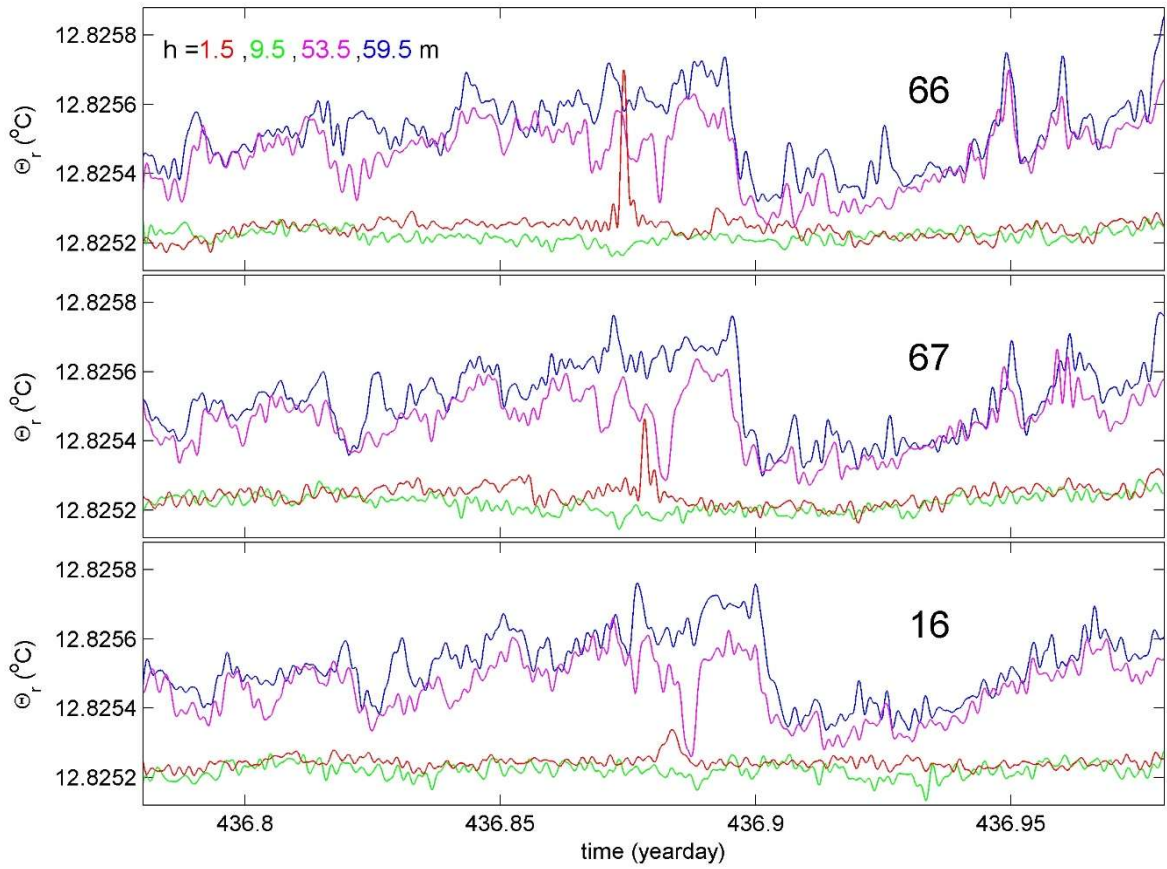

**Figure 5.** Enlargements to 0.2-day time series of three panels of Figs 3, 4 but for lpf Conservative Temperature referenced to 124-m tall mean vertical profile corrected to a 3rd-order polynomial fit.

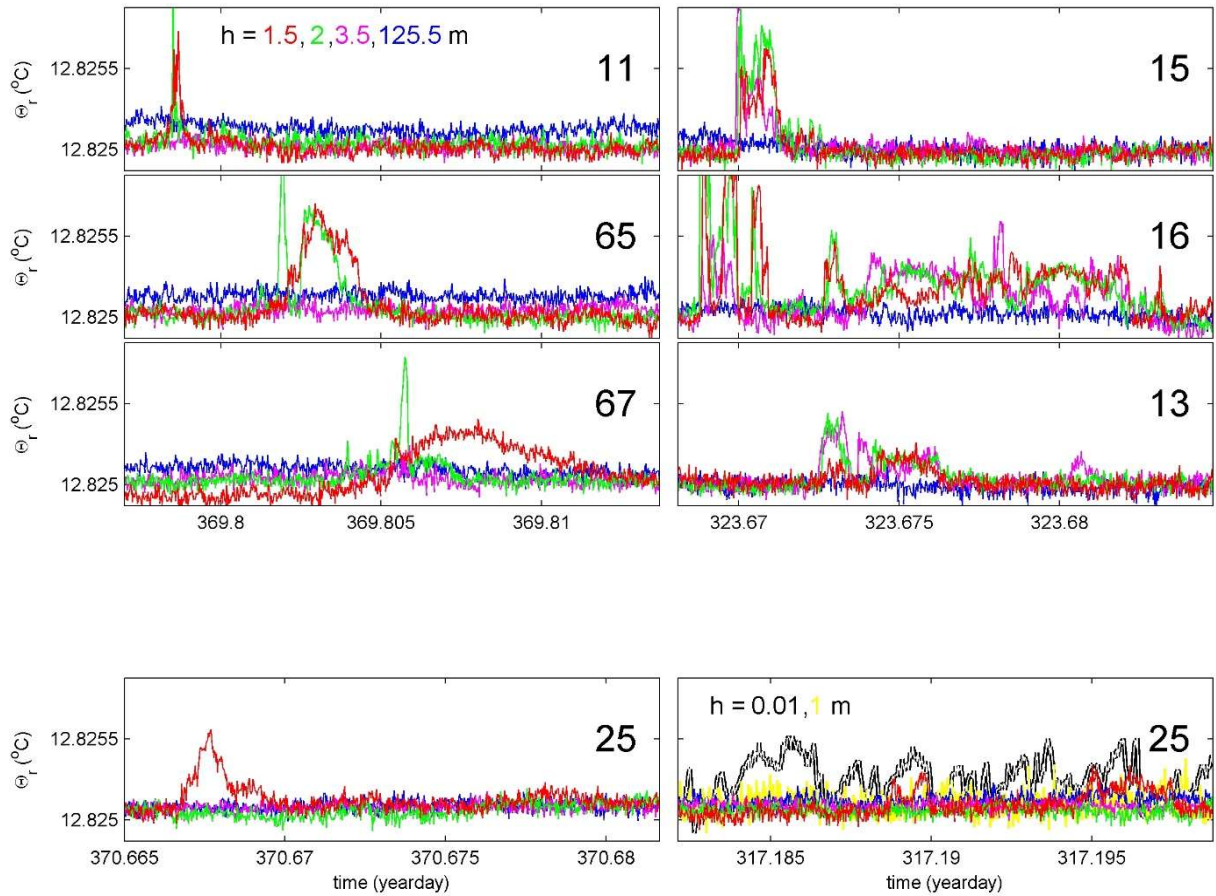

**Figure 6.** As Fig. 3, but for a glossary of shorter horizontal-range examples, whereby the uppermost T-sensor replaces that of the fourth from the seafloor. Upper-three left column: a sequence of heat flashes akin to Fig. 3 but traveling to the WNW during an episode of very weak stratification. Lower left: singular isolated flash at the lowest sensor of line 25. Upper-three right column: Remotely Operated Vehicle 'ROV' working to remove parachute from nearby line 18. Lower right: multiple flashes at lower sensor of line 25, also registered at line 16, in comparison with short-term sensors closer to the seafloor at acoustic release (yellow) and line-18 fallen-off sensor (black&white).

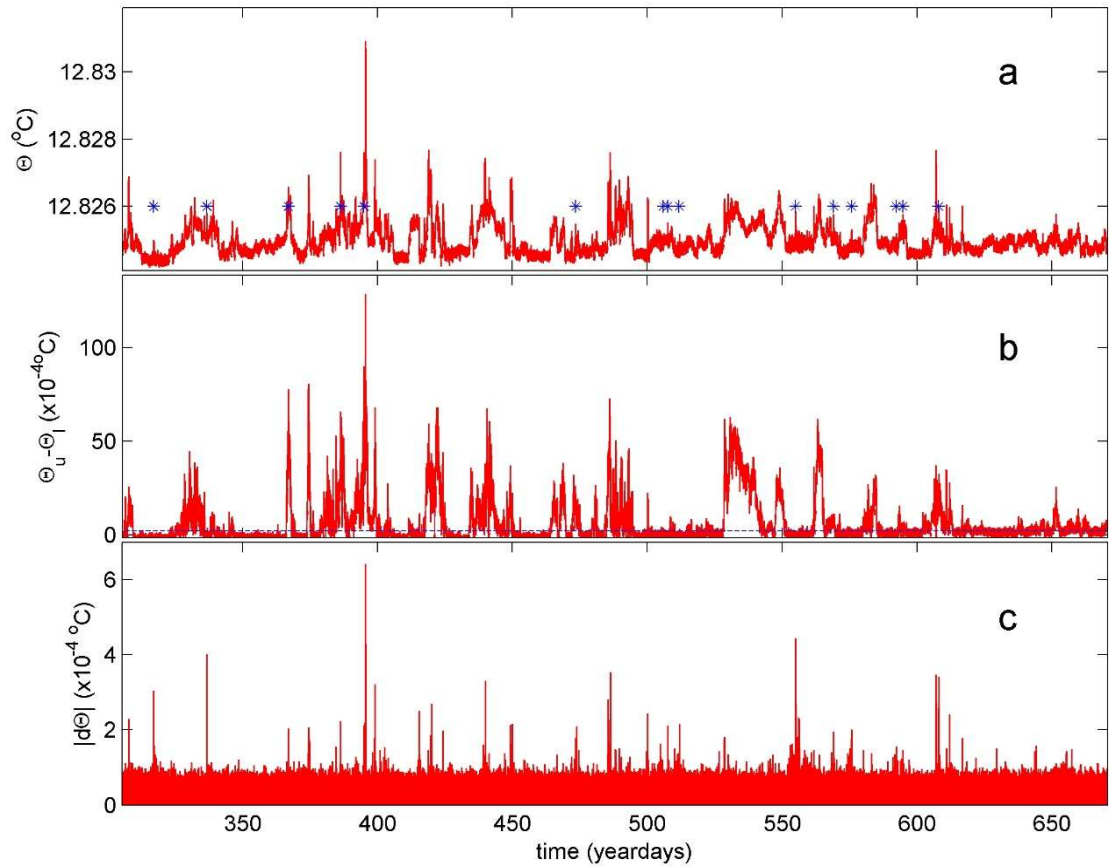

**Figure 7.** Year-long time series indicating occurrence of heat flashes at lowest T-sensor of corner-line 17. Temperature data are not corrected for bias. (a) Conservative Temperature with stars indicating heat flashes. (b) Upper-lowermost T-sensor's 124-m vertical temperature difference with dashed line indicating a threshold of 0.0002°C, approximately designating near-homogeneous from stratified-water conditions. (c) Gradient (with time) of record in a.

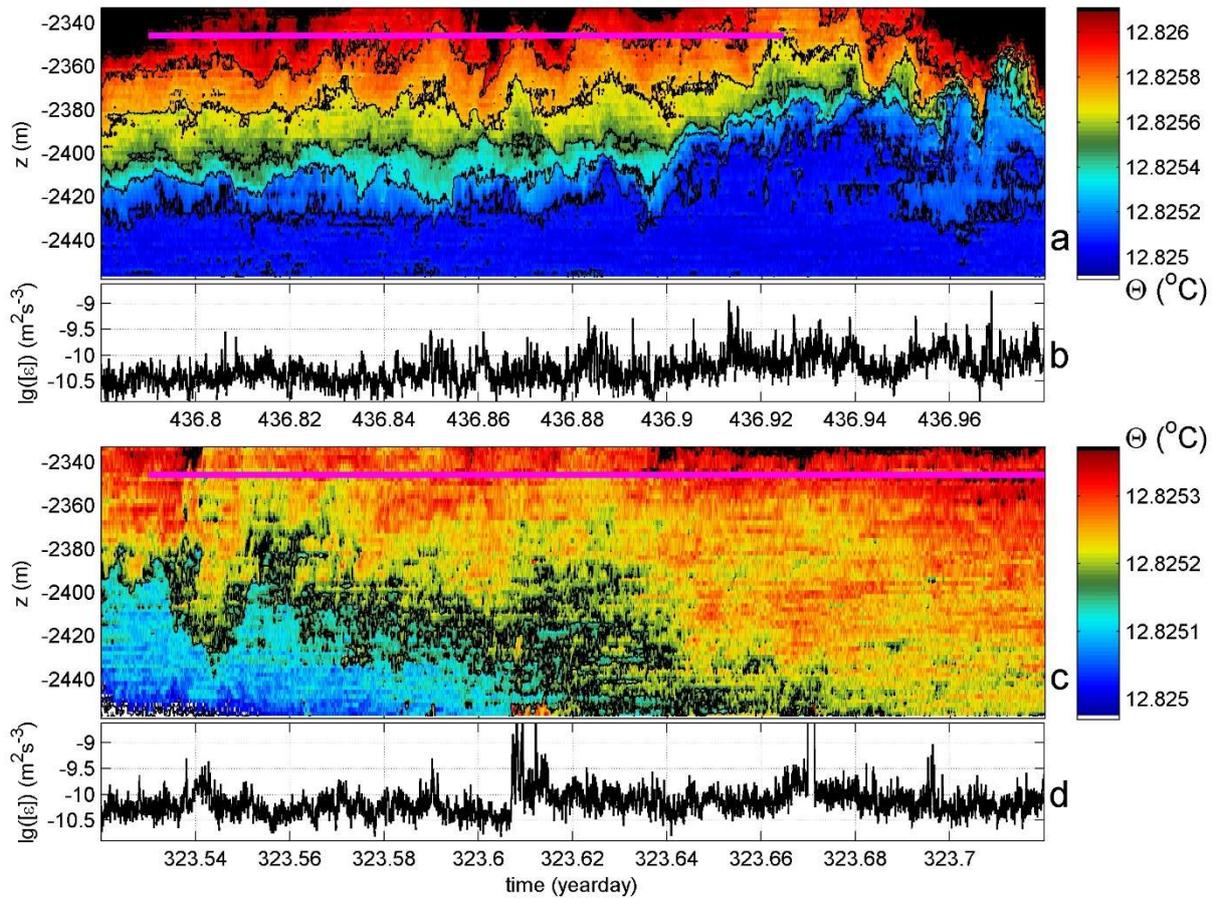

**Figure 8.** Magnifications of 0.2-day time and 124-m depth images with vertically averaged turbulence dissipation rate values using the method by Thorpe (1977). (a) Conservative Temperature from line 16 lpf at cut-off frequency of 3000 cpd (short for cycles per day) for weakly-stratified-water period encompassing that of Figs 2, 3. Black contours are drawn every 0.0002°C. The magenta bar demonstrates the duration of the shortest free internal-wave period during the first 40% of data. The seafloor is at the horizontal axis. The small speckles just above the seafloor denote heat flashes, the largest passing at day 436.882. (b) Logarithm of corresponding 124-m vertically averaged dissipation rates. (c) As a., but for line-15 data during very weakly stratified period (three-times smaller temperature range than in a.) and ROV working near line 18, mainly between days 323.605 and 323.67. The magenta bar extends out of window, covering about 0.22 d. (d) As b., corresponding to c.

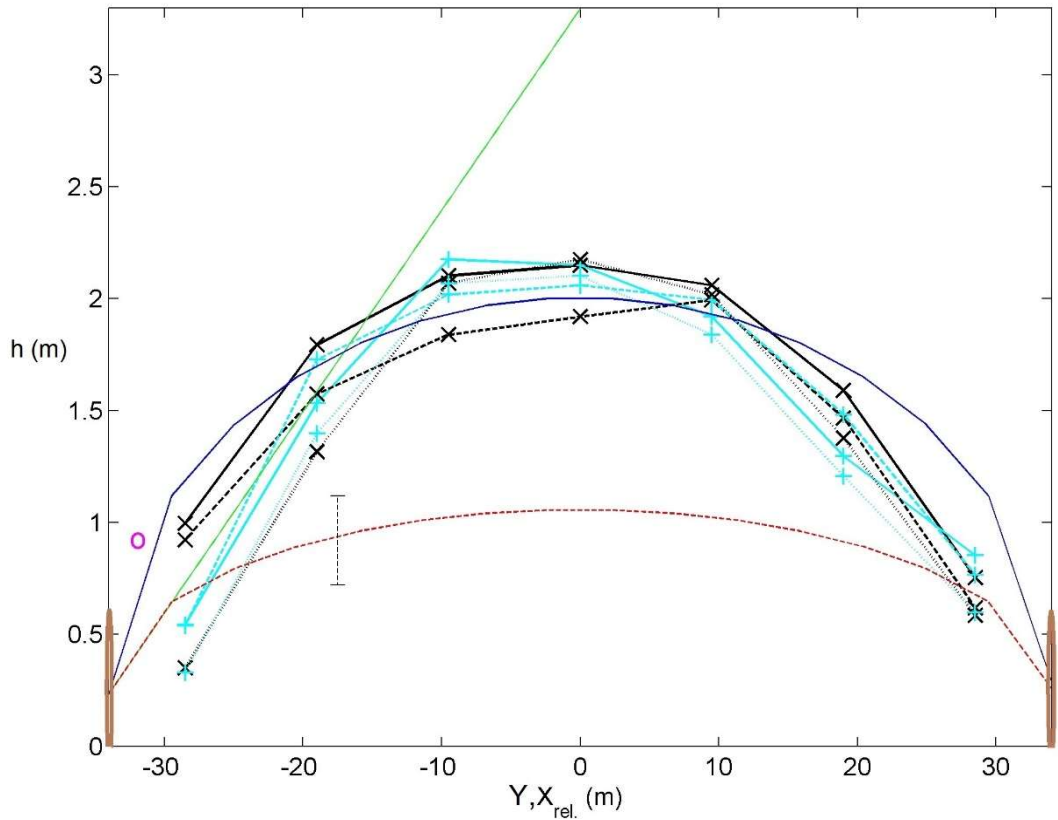

**Figure A1.** The height doming of the large-ring mooring steel-cable grid under tension of 45 buoys as followed from modelling and observations by van Haren (2026b; submitted). Quasi-parabola (red, blue) and straight-line (green) mathematical models are given for center cables. The seafloor is at the horizontal axis. Cable-grid attachments are halfway large-ring pipes (brown ellipses), effectively at h = 0.24 m above seafloor as the anchoring pipes sunk 0.07±0.02 m in the sediment. The green straight line makes a fixed angle of 5° with the horizontal, which angle was established after in-port cable-tension tests. The blue (solid line, 5-m discretized) parabola model intersects the green line halfway, so that its top is at h = 2.00 m. If an overall maximum angle of 5° is maintained (red-dotted model), the top is at h = 1.05 m, and the lowest vertical line will be at h = 0.65 m. Cross-sections without corner-lines of constant-Y (black x graphs) and constant-X (cyan + graphs) are given for physical relative heights determined from observed turbulence temperature-variance. Solid lines indicate center lines in both directions. Corner-line height is indicated by magenta o.